\documentclass[twoside]{article}

\usepackage{PRIMEarxiv}

\usepackage[utf8]{inputenc} 
\usepackage[T1]{fontenc}    
\PassOptionsToPackage{hyphens}{url}
\usepackage{hyperref}       
\usepackage{url}            
\usepackage{booktabs}       
\usepackage{amsfonts}       
\usepackage{nicefrac}       
\usepackage{microtype}      
\usepackage{lipsum}
\usepackage{fancyhdr}       
\usepackage{graphicx}       
\usepackage[round]{natbib}

\title{Automated clustering of COVID-19 anti-vaccine discourse on Twitter
\thanks{\textit{\underline{Correspondence}}:
 \texttt{ignacio.ojeaquintana@anu.edu.au},    \texttt{marc.cheong@unimelb.edu.au}} 
}

\author{
  Ignacio Ojea Quintana$^{*}$ \\
  Department of Philosophy\\
  Australian National University\\
  ACT Australia \\

  \And
  Marc Cheong$^{*}$  \\
  Centre for AI and Digital Ethics \& \\
  School of Computing and Information Systems\\
  University of Melbourne\\
  VIC Australia \\
  
  \And
  Mark Alfano \\
  Department of Philosophy\\
  Macquarie University\\
  NSW Australia \\
  
  \And
  Ritsaart Reimann\\
  Department of Philosophy\\
  Macquarie University\\
  NSW Australia \\
  
  \And
  Colin Klein\\
  Department of Philosophy\\
  Australian National University\\
  ACT Australia \\

}

\begin{document}

\maketitle

\begin{abstract}
Attitudes about vaccination have become more polarized; it is common to see vaccine disinformation and fringe conspiracy theories online. An observational study of Twitter vaccine discourse is found in
\cite{PLOSMANUSCRIPT}: the authors analyzed approximately six months’ of Twitter discourse — 1.3 million original tweets and 18 million retweets between December 2019 and June 2020, ranging from before to after the establishment of Covid-19 as a pandemic.

This work expands upon \cite{PLOSMANUSCRIPT} with two  main contributions from data science. First, based on the authors' initial network clustering and qualitative analysis techniques, we are able to clearly demarcate and visualize the language patterns used in discourse by \textit{Antivaxxers} (anti-vaccination campaigners and vaccine deniers) versus other clusters (collectively, \textit{Others}). Second, using the characteristics of Antivaxxers' tweets, we develop text classifiers to determine the likelihood a given user is employing anti-vaccination language, ultimately contributing to an early-warning mechanism to improve the health of our epistemic environment and bolster (and not hinder) public health initiatives.

\end{abstract}

\section{Introduction}
\label{section:Introduction}

Vaccine distrust, disinformation, and hesitancy, is hampering the rollout of vaccines, especially in the COVID-19 crisis, as herd immunity is seen as the way out of the pandemic. Sadly, almost three years into the pandemic (as of time of writing), the issues are far from resolved.

Hence, anti-vaccine campaigners on social media-- or \textit{Antivaxxers} -- ``spreading misinformation that undermines the rollout of vaccines against COVID-19'' \citep{Ahmed2021}, are a cause for alarm, as they are concerted, ``well financed, determined and disciplined'' \citep{Ahmed2021}.

The impetus for this paper is \cite{PLOSMANUSCRIPT}, an extensive study of data from Twitter during the early onset of the COVID-19 pandemic worldwide to determine the dynamics of vaccine discourse, with an emphasis on (dis)trust of expertise \citep{NguyenForthcoming,Yaqub2014}. To wit, ``a trust-first dynamic of political engagement... [with a subsequent] realignment of interests over the course of the pandemic'' \citep{PLOSMANUSCRIPT} was studied using techniques such as natural language processing, Linguistic Inquiry and Word Count (LIWC)-based approaches and vector autoregression.

In this paper, we set out to extend \cite{PLOSMANUSCRIPT} using a further combination of methods -- observational analysis as well as standard data science classification methods -- to answer the following questions:

\begin{itemize}
    \item \textbf{RQ1} What are emergent characteristics of the discourse in the Antivaxx cluster versus all the other clusters (even accounting for political rhetoric), which are manifest in the writing style of their tweets?
    \item \textbf{RQ2}. How can we build a simple binary classifier -- in the spirit of, say, spam and phishing detectors on email -- to separate between vaccine information and misinformation, learning from the above emergent characteristics?
\end{itemize}

This paper is organised as follows: \textit{Related Work} covers extant research related to each of our \textbf{RQ}s; \textit{Methods} details our approaches; leading into an analysis of our results in \textit{Results} before lastly finishing with a review and recap in \textit{Discussion and Conclusion}. 

\section{Related Work}
\label{section:RelatedWork}

In this section, we briefly touch on related work relevant to our investigation.

\subsection{Network Analysis and Social Media}
Network and cluster analysis are the norm in studies on social media, from both the network perspective as well as the messages shared. Retweet networks have been commonly used in the past few decades of Twitter research \citep{Boyd2010,Zhu2016,Cheong2013,Bovet2019,Sullivan2020a,Sullivan2020b}. \cite{Pacheco2020} have conducted recent work on `coordination on social media', identifying that `spatiotemporal patterns of activity' and `content being shared' can intertwine in a concerted effort to coordinate a particular message online: they focused on the networked structure of users based on re-shared images and retweets and hashtags. Even pre-COVID times, polarization is found in English-language discourse about vaccines on Twitter, when it comes to issues such as vaccine safety \citep{Sullivan2020a,Sullivan2020b}.


A recent study by \cite{An2020} found the effects of \textit{homophily} -- the propensity of a user interacting with other users with similar properties -- when analyzing political tweets concerning US presidential candidates  \citep{McPherson2001}. They conducted a ``simulation of six different Twitter accounts representing archetypes of heavily active and moderately activity Republican and Democrat-leaning panel members, as well as control accounts representing individuals who might use Twitter very sparingly'' \citep{An2020}. Based on the intertwining between responses to COVID-19 and political affiliation we raised earlier, we can infer that one who adopts a certain worldview about vaccines will exhibit the same homophilic behaviour in their social network, worthy of examination.

\subsection{COVID-19 and Language on Social Media}
The COVID-19 \textit{infodemic} -- a term used by the WHO to describe ``an overabundance of information, both online and offline... [which] includes deliberate attempts to disseminate wrong information to undermine the public health response and advance alternative agendas'' \citep{World_Health_Organization2020} -- is characterised as, amongst others, ``...spreading fake news, rumors, and conspiracy theories, and extends to promote fake cures, panic, racism, xenophobia, and mistrust in ... authorities'' \citep{Alam2020}. \cite{Alam2020} has contributed a thorough literature review on research on fake news/misinformation and COVID-19 social media activity.

For Twitter language analysis, we turn to  \cite{Reiter-Haas2021} who examined `differences in moral framing' of politicians' use of Twitter: more specifically, COVID-19 related tweets from Austrian political parties. Parts of their work are based on Moral Foundations Theory (MFT), a framework consisting of domains or ``foundations'' \citep{Graham2009,Graham2013}, by analyzing words corresponding to these topics (care/fairness/loyalty/sanctity).
An alternative to MFT, which is the Morality-as-Cooperation (MAC) hypothesis \citep{Curry2019a}, also has potential for investigations in this area of interest. 

\subsubsection*{Automated Binary Classification}
Existing studies on, say, binary spam classification on email text -- ranging from traditional techniques \citep{USENIX-SPAM} to more modern ones \citep{DADA2019e01802} -- are also worth mentioning when it comes to our main \textbf{RQ}s, in terms of best practices and the state of the art.



\section{Methods}
\label{section:Methods}

In this section, we explain the methods used to collect, clean, and curate the dataset, based on \cite{PLOSMANUSCRIPT}; the methods of inquiry to investigate our \textbf{RQ}s; as well as technical and philosophical considerations in our analysis.

\subsection{Data Collection}

\subsubsection{Twitter API Preliminaries}
The Twitter Streaming API was queried with a series of vaccination-related terms and hashtags (date range: December 2019 -- June 2020), per \cite{PLOSMANUSCRIPT}.\footnote{The following pre-registered vaccination-related keywords, hashtags and short expressions were used: \textit{'vax', 'vaxxed', 'vaccine', 'vaccination', 'vaccinations', 'vaxsafety', 'vaccineswork', 'vaccines work', 'vaccines revealed', 'vaccinesrevealed', 'novax', 'no vax', 'no-vax', 'antivax', 'anti-vax', 'anti vax', 'immunisation', 'Vaccin', 'Vaccinaties', 'vaccinatiezorg', 'vaccine injury', 'vax injury','vaccinatieschade', '\#vax', '\#vaxxed', '\#vaccine', '\#vaccination', '\#vaccinations', '\#vaxsafety', '\#vaccineswork', \#vaccinesrevealed', '\#novax', '\#antivax', '\#immunisation', '\#Vaccin', '\#Vaccinaties', '\#vaccinatiezorg', '\#vaccinatieschade', '\#nvkp', '\#rvp', '\#rijksvaccinatieprogramma', '\#vaccineinjury', '\#vaxinjury', '\#anti-vax'}}

 We considered only retweets in our analysis (not quote tweets), hereinafter termed \textit{posts}, as retweets are considered endorsements of their original posts (see also \cite{Cheong2013} and \cite{Boyd2010}).

\subsubsection{Bot Activity and Inclusion Criteria}
For this paper, we decided to include any posts that may have been generated by bots (e.g. spam bots, or those involved in a mass disinformation campaign). This design decision was made as we wanted to observe the effects of such bots within the overall discourse. Moreover, and given our ambition to train a classifier capable of detecting the spread of misinformation (\textbf{RQ2}): removing bot generated text would hamper the accuracy of any proposed classifier as it would inherently lack the ability to classify misinformation generated by bots.

For completeness, we did run a cursory check for bot activity by examining the top 50 authors by post and by retweet count. Upon manual inspection, we found that most prominent accounts do not behave like bots. To mitigate the noise of bot activity, users that only retweeted (i.e. without other contributions) were disregarded.

\subsubsection{Network Construction and Community Detection}
As per \cite{PLOSMANUSCRIPT}, we implemented the network, and subsequent community detection as follows:
\begin{description}
    \item[constructing a network:] a weighted directed network where nodes are authors and the weight of an edge $u \rightarrow v$ represents the number of times that user $v$ retweeted user $u$ \citep{PLOSMANUSCRIPT};
    \item[discarding any self-retweeting:] as the intended interpretation of the network is that $u \rightarrow v$ implies that `$u$ influenced $v$';
    \item[pruning:] only the principal weakly connected component is used; 231 communities were obtained, with the top five largest considered (comprising $\sim$80\% of the population) \citep{PLOSMANUSCRIPT}; 
    \item[characterizing communities:] we considered the top verified accounts and hashtags used within each of the clusters (communities). 
\end{description}

\subsubsection{Addressing RQ1: Characteristic Linguistic Styles}
To address \textbf{RQ1}, we need to analyse the characteristic linguistic styles of each community/group; most importantly, differentiating between Antivaxxers' use of language and non-Antivaxxers' linguistic patterns (the latter which may include politically-charged language as well). 

An exploratory analysis of the most characteristic terms used by Antivaxxers as opposed to non-Antivaxxers will be conducted using the Scattertext \citep{kessler2017scattertext} package in Python, which gives us an idea of the language used by those advocating vaccine hesitation, distrust, and denialism, when compared to other groups.

This analysis serves to complement the manual analysis done in prior work \citep{PLOSMANUSCRIPT}.

\subsection{Addressing RQ2: Classifier Construction}
To answer \textbf{RQ2}, we use Python to construct a workflow for automated text classification; in particular Tensorflow \citep{tensorflow2015-whitepaper} and Scikit-Learn \citep{scikit-learn}. As we are interested to distinguish between language used by Antivaxxers versus other communities on Twitter (again controlling for e.g. political rhetoric), we set it up as a pure \textit{binary classification task}, in contrast to \cite{PLOSMANUSCRIPT}'s use of determining cluster membership.

The pipeline is as follows. For each user in our network we define a \textit{document} for said user as \{the plain text of tweets they authored + the emojis + hashtags used\}. Tweet text was pre-processed (e.g., removing non alpha-numeric characters, standardizing case), and emojis were encoded using Python's \textit{emoji} package \citep{emoji}. 

With respect to emoji, recent work suggests that they are increasingly used to convey sentiments and concepts in a graphical manner \citep{Robertson}. Furthermore, distinguishing between the three linguistic tools (text, hashtag, emojis) can be useful for further analysis, although we decided to bundle them together for this project. \#hashtags are traditionally used on Twitter to classify messages  \citep{Cheong2013,Sullivan2020a}. 

We constructed our training and validation data sets in the following manner\footnote{See also \citep{BLMMANUSCRIPT} for similar techniques}:
\begin{description}
\item[Sampling:] all the data corresponding to users in the Antivaxx community ($\sim$ 30K authors - positive instances) and a random sample of users in other communities (negative instances) are used for a 50/50 split.
\item[Train/Test Split:] a 80/20 train/test split is used.
\end{description}
Our results are stable for both.

We used two different data representations and corresponding classifiers. 
\begin{description}
\item[Bag-of-words:] for classical ML techniques in Scikit-Learn (logistic regression, random forests, naive Bayes, etc.): tokens (words + emoji + hashtags) are features.
\item[Sequential representation:] deep learning techniques (Tensorflow DNNs, GRUs, LSTMs); Succinctly, "each document is now a sequence, where each word token is encoded as a number. Later, in learning, those numbers are embedded into a vector, so that each document is effectively a matrix" \citep{BLMMANUSCRIPT}
\end{description}

For evaluation, we used \textit{accuracy} and \textit{Area Under the [ROC] Curve (AUC)}. 

\section{Results}
\label{section:Results}

In this section, we report our results, beginning with community characterization and corpus analysis.

\subsection{Preliminaries}
\label{section:communitycharacterization}

\begin{figure}[!htb]
\centering
\includegraphics[width=0.75\textwidth]{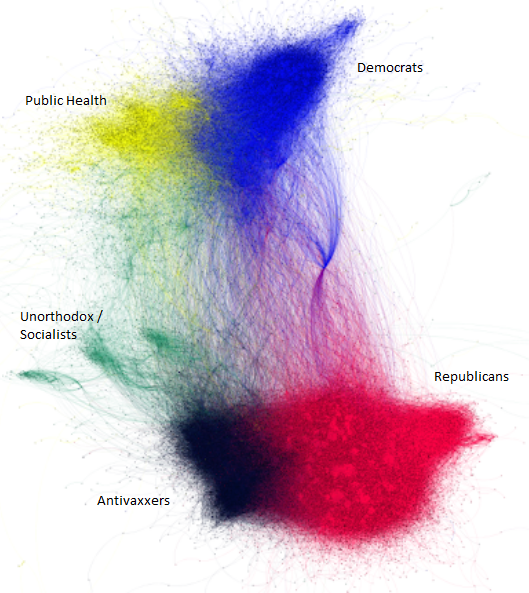}
\caption{Visualisation of the retweet network, color-coded by community. Original image in \cite{PLOSMANUSCRIPT}.}
\label{fig:fullnetwork}
\end{figure}

Before reporting on \textbf{RQ1}, it is pertinent to note the existing clustering of the 5 communities per \cite{PLOSMANUSCRIPT}. These communities are clearly interpretable (Figure \ref{fig:fullnetwork}), containing $\sim80$\% of the nodes and are responsible for $\sim90$\% of the retweets. For this study, we decided to focus on the properties of the top five communities. Similar to \cite{PLOSMANUSCRIPT}, Table \ref{table:communities} summarizes the emergent features of each of our named communities.

\begin{table*}[htb]
\centering
\caption{Summary statistics for the five communities analyzed in this study. Partially adapted, and expanded, from \cite{PLOSMANUSCRIPT}.}

\begin{tabular}{|l|c|c|p{0.25\textwidth}|} 
\hline
Community name & Proportion of nodes & \% of retweets & Description \\
\hline\hline
Democrats & $\sim24$\% & $\sim20$\% & Democratic politicians
and some center-left media.\\
\hline
Republicans & $\sim18$\% & $\sim35$\% & Republican politicians
and some right-wing media.\\
\hline
Unorthodox & $\sim16$\% & $\sim6$\% & Non-mainstream activists which include socialists.\\
\hline
Public Health & $\sim13$\% & $\sim7$\% & Public health experts and organizations such as the CDC, UNICEF, and academics/universities.\\
\hline
Antivaxxers & $\sim8$\% & $\sim22$\% & Vaccine deniers, which includes conspiracy theorists. \\
\hline  
\end{tabular}
\label{table:communities}
\end{table*}

Of interest are the top hashtags used by Antivaxxers versus the rest of the communities. They include hashtags such as \textit{\#illuminati, \#praybig, \#notest, \#mykidsmychoice, \#vaccineroulette}: these all display signs of vaccine hesitancy and skepticism. All other communities include political rhetoric\footnote{With the exception of \textit{\#illuminati} signifying a conspiracy theory which is not in the Antivaxxer cluster.}, scientific discussions, and activism campaigns.







\subsection{RQ1: Detecting Antivaxxer Rhetoric}
Having a better sense of the communities involved in the vaccine discourse \citep{PLOSMANUSCRIPT}, we now address \textbf{RQ1} by comparing the language patterns used in the Antivaxx cluster with the patterns found in other communities. 

The strategy of comparing one-versus-the-others is useful as we aim to distinguish between Antivaxxer rhetoric and the rhetoric of the other communities, which, presumably \textit{do not produce or promote}  disinformation and misinformation. To elaborate, any posts in support of (or attacking) a given political party (Democrats/Republicans), in support of (or refuting) `unorthodox' activists, and those coming from scientists and health professionals, are treated as \textit{Others} in this respect.

To explore the gamut of words and phrases used by Antivaxxers and Others, we employ Scattertext \citep{kessler2017scattertext} to visualise the distribution of words and bigrams that occur at least 5,000 times in the entire corpus. The default options are otherwise used when invoking the Scattertext library.

\begin{figure*}[!htb]
\centering
\includegraphics[width=1.0\textwidth]{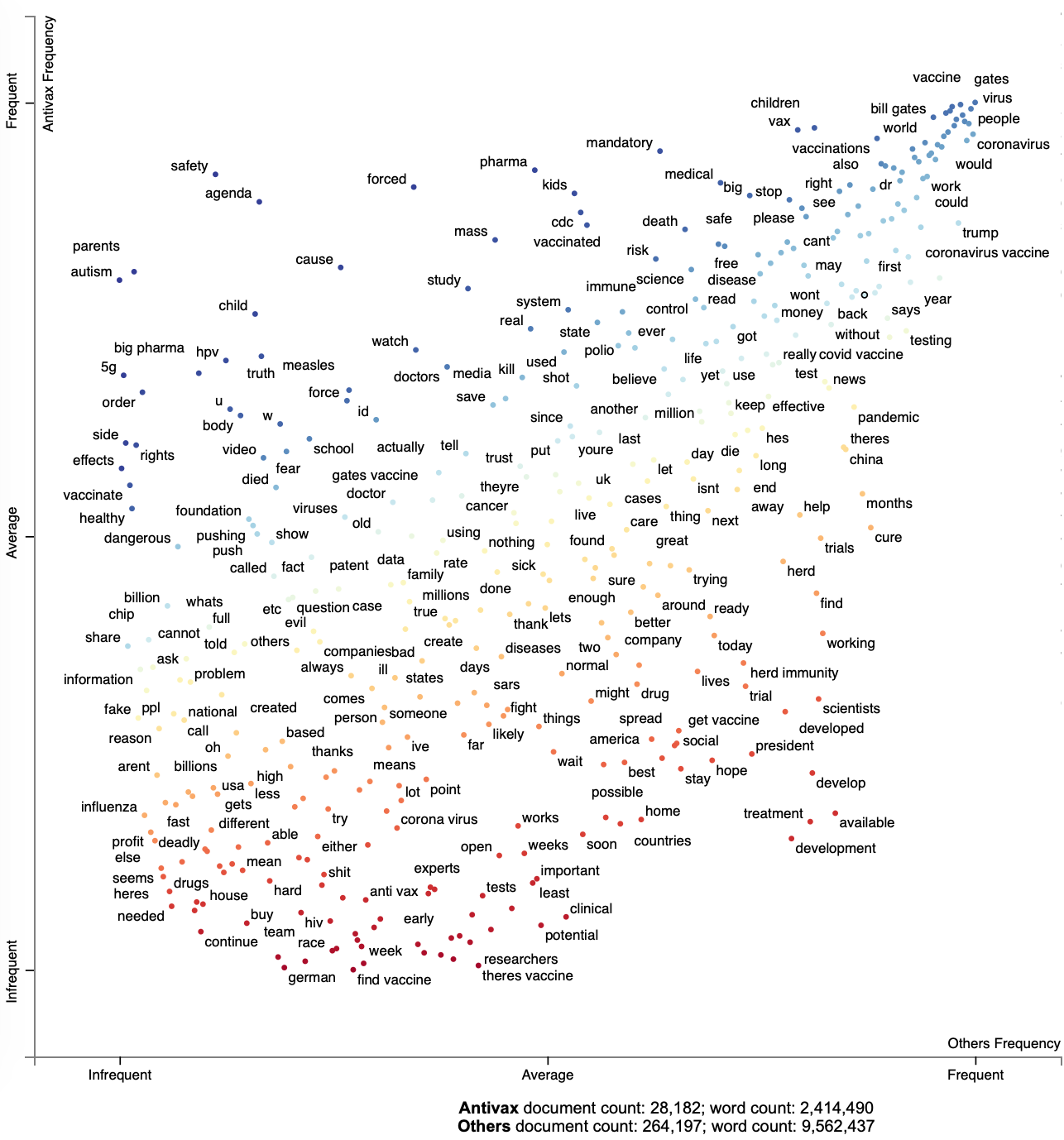}
\caption{Scattertext distribution of words and bigrams occurring at least 5,000 times in our corpus, minus stopwords. Words which are not in the English language dictionary are omitted. The likelihood of a word/bigram appearing in Antivaxxers posts are based on its $x$-axis position (left: low/right: high), and the likelihood of it appearing in non-Antivaxxer clusters (\textit{Others}, per the figure)  are based on the $y$-axis position (bottom: low/top: high).}
\label{fig:scattertext}
\end{figure*}

In Figure \ref{fig:scattertext}, the spatial properties of points (words/bigrams) in the Scattertext visualization are based on likelihood of said point occurring in both Antivaxxers posts ($x$-axis) versus Other (non-Antivaxxer) communities' posts ($y$-axis). 

Words that are commonly associated with vaccine denial and hesitation -- such as \textit{autism}, \textit{big pharma}, \textit{truth}, and \textit{5g}\footnote{Indicative of  COVID-19 conspiracy theories about mobile networks.} are in the top-left quadrant, indicating a greater-than-usual prevalence in the Antivaxx cluster. Words in the bottom-right quadrant (i.e. predominantly non-Antivaxx) are scientific and research-oriented in nature, though political rhetoric becomes more prevalent as we move closer to the centre diagonal. Interestingly, mentions of former US president \textit{Trump}, as well as \textit{Bill Gates}\footnote{Indicative of another COVID-19 conspiracy theory.} are common across both Antivaxxers and non-Antivaxxers.

\subsection{RQ2: Does A Simple Binary Classifier Work?}

 The results for the classification task are
 shown in Table \ref{table:classification} below.
 
\begin{table*}[htb]
\centering
\caption{Classification task results}

\begin{tabular}{|l|c|c|c|p{0.25\textwidth}|p{0.25\textwidth}|} 
\hline
Classifier & Accuracy & AUC & Data Representation  \\
\hline\hline
Logistic Regression & 0.73 & 0.81 & Bag of Words \\ \hline
Random Forest & 0.74 & 0.81 & Bag of Words \\
\hline
Linear SGD & 0.68 & 0.74 & Bag of Words \\
\hline
Multinomial NB & 0.57 & 0.74 & Bag of Words \\ \hline
DNN & 0.75 & 0.83 & Bag of Words \\ \hline
DNN & 0.75 & 0.83 & Sequential \\ \hline
GRU & 0.76 & 0.84 & Sequential \\ \hline
LSTM & 0.77 & 0.85 & Sequential \\

\hline

\hline  
\end{tabular}
\label{table:classification}
\end{table*}

The best performing classifier was an LSTM neural network, with an average accuracy of about 77\% and AUC of .85. Naive Bayes classifiers were the worst performers, the best of them reaching 57\% accuracy and 0.74 AUC. 

Three observations are in order. First, deep learning techniques marginally outperformed more classical approaches. Second, there was no difference between representation structures (bag of words vs sequential) for standard DNN. Finally, GRU and LSTM marginally outperfomed DNN, suggesting that there is something to learn from the ordering.

We answered \textbf{RQ2} by making an inference from retweet behavior to linguistic behavior. In other words, the network encodes the retweet behavior of individual users during a period of time. Using an unsupervised technique, namely modularity clustering, we identified and described different communities. We later used these labelings, that were generated without supervision, and employed supervised techniques (i.e. classifiers) to identify linguistic behavior (i.e. antivaxx discourse). It is unsurprising that retweet behavior and linguistic behavior are connected. We are exploiting that fact in order to answer \textbf{RQ2}.

\section{Discussion and Conclusion}
From our investigation above, we have satisfactorily explored the properties of language used by Antivaxxers in their social media posts, in comparison to other users on social media, which are categorized/clustered into communities based on retweet behavior.

Firstly, we were able to identify commonalities between users motivations for Twitter use during the COVID-19 pandemic, when the topic of discourse centred around vaccines, especially to single out users who are part of the Antivaxx movement. We visualized the landscape of language use by Antivaxxers versus other clusters/communities, to determine common themes found within. 
This allows us to also validate our results based on published observations on vaccine hesitancy and denialism.

The task in \textbf{RQ2} was to classify and identify text that contain antivaxx features and messages. This can be used to stem the COVID-19 online infodemic; and consequently supports and enables efforts to promote correct public health outcomes and encourage vaccination for herd immunity. As explained before, we answered \textbf{RQ2} by making an inference from retweet behavior (network structure) to linguistic behavior (Twitter authors' corpus of published tweets).

Recognizing that our work is a pilot study, we note down several points where this work can be improved in future iterations. 

In addressing said  \textbf{RQ}s, future work would involve a closer cooperation with experts in public health messaging to identify the nuances in the Antivaxxers messaging strategy, and to provide effective psychological `inoculation' \citep{Ahmed2021} against such mis-/disinformation campaigns online. Examples may include \textit{nudging} and \textit{boosting}, which have been used to thwart microtargeting strategies online \citep{Lorenz-Spreen2020}.

An astute reader might point out that classifiers are better suited at the tweet level than the author level; namely classifing antivaxxer \textit{tweets} rather than \textit{authors}. Although we leave that task for future work, there were two reasons for the alternative in this essay. First, an author's corpus can be studied across platforms, and therefore our classifiers could be of more use for other researchers. Second, the network was generated studying the aggregate information about their retweet behavior during a period of time, it is more natural to study the correlation with the author's corpus during that time; in contrast individual tweets are dated, non-aggregated, linguistic actions.

A further limitation is that our classifiers are somewhat coarse-grained and may thus \textit{cast a wide net}. Which is to say, we cannot guarantee that authentic and well-intentioned users who are tweeting and sharing accurate content will not be flagged as malicious misinformers. Any judgment on particular individuals ought to be made by taking onto consideration variables not studied here.

For \textbf{RQ2}, we have also identified that emojis are an under-researched area when it comes to social media: the use of emojis in signalling concepts, emotions, and influencing behaviour deserves a more thorough exposition. Further work in this area would include a more thorough feature engineering with emojis to improve classifier accuracy.

\section*{Data Access}
Data representations, classifiers, as well as neural networks weights for transfer learning are available upon request from the corresponding authors.

\bibliographystyle{plainnat}
\bibliography{main}

\end{document}